\documentclass[aoas,MSNbibl,nameyear,dvips]{arximspdf}
\usepackage{dcolumn}
\usepackage{graphicx}

\doi{10.1214/11-AOAS517}
\volume{6}
\issue{2}
\pubyear{2012}
\firstpage{476}
\lastpage{496}

\makeatletter
\newtheorem{thmm}{Theorem}
\newproclaim{defn}{Definition}
\newcolumntype{d}[1]{D{.}{.}{#1}}
\newcommand{\eqref}[1]{(\ref{#1})}
\renewcommand{\citep}[1]{\citeauthor{#1} \citeyear{#1}}
\makeatother

\begin{document}
\begin{frontmatter}

\title{Change-point model on nonhomogeneous Poisson processes with
application in copy number profiling by next-generation DNA
sequencing\thanksref{T1}}
\runtitle{Change-point model and copy number with sequencing}

\begin{aug}
\author[A]{\fnms{Jeremy J.} \snm{Shen}\corref{}\ead[label=e1]{jqshen@stanford.edu}}
\and
\author[B]{\fnms{Nancy R.} \snm{Zhang}\ead[label=e2]{nzhang@stanford.edu}}
\runauthor{J. J. Shen and N. R. Zhang}
\affiliation{Stanford University}
\thankstext{T1}{Supported by NSF Grants DMS-10-43204 and DMS-09-06394.}
\address[A]{Department of Statistics\\
Stanford University\\
Sequoia Hall, 390 Serra Mall\\
Stanford, California 94305-4065\\
USA\\
\printead{e1}} 
\address[B]{Department of Statistics\\
The Wharton School\\
University of Pennsylvania\\
3730 Walnut Street, Suite 467\\
Philadelphia, Pennsylvania 19104\\
USA\\
\printead{e2}}
\end{aug}

\received{\smonth{10} \syear{2010}}
\revised{\smonth{9} \syear{2011}}

%
\begin{abstract}
We propose a flexible change-point model for inhomogeneous Poisson
Processes, which arise naturally from next-generation DNA sequencing,
and derive score and generalized likelihood statistics for shifts in
intensity functions. We construct a modified Bayesian information
criterion (mBIC) to guide model selection, and point-wise approximate
Bayesian confidence intervals for assessing the confidence in the
segmentation. The model is applied to DNA Copy Number profiling with
sequencing data and evaluated on simulated spike-in and real data sets.
\end{abstract}

%

\end{frontmatter}

\section{Introduction}\label{sec1}

For a biological sample, the DNA copy number of a~genomic region is
the number of copies of the DNA in that region within the genome of
the sample, relative to either a single control sample or a~pool of
population reference samples. DNA Copy Number Variants (CNVs) are
genomic regions where copy number differs among individuals.
Such variation in copy number constitutes a common type of population-level
genetic polymorphism. See \citet{Khaja}, \citet{Redon}, \citet{conrad} and
\citet{McCarroll} for detailed discussions on CNV in the human population.

On another front, the genomes of tumor cells often undergo somatic
structural mutations such as deletions and duplications that affect
copy number. This results in copy number differences between tumor cells
and normal cells within the same individual. These changes are often
termed Copy Number Aberrations or Copy Number Alternations (CNA).
There is significant scientific interest in finding CNVs in normal
individuals and CNAs in tumors, both of
which entail locating the boundaries of the regions in the genome that
have undergone copy number change
(i.e., the breakpoints), and estimating the copy numbers within
these regions. In this article, we use next-generation sequencing
data for copy number estimation.

Microarrays have become a commonly used platform for high-throughput measurement
of copy number. There are many computational methods that estimate
copy number using the relative amount of DNA hybridization to an array.
See \citet{LaiPark}, \citet{Willenbrock2005} and \citet{Zhang2010CNBookChapter}
for a general review of existing methods for array-based
data. However, the precision of breakpoint estimates with array-based
technology is
limited by its ability to measure genomic distances between probes,
which currently
averages about 1000 bases (1~Kb) on most arrays. Hence, the lower limit
in the length
of detectable CNV events is about 1 Kb. With sequencing capacity
growing and its cost dropping dramatically, massively
parallel sequencing is now an appealing method for measuring DNA
copy number. In these newer sequencing technologies,
a large number of short reads (36--100 bp) are sequenced in parallel
from the fragmentation of sample DNA. Then each read is mapped to
a reference genome. The basic rationale is that \textit{coverage},
defined as the number of reads mapped to a region of the reference
genome, reflects
the copy number of that region in the sample, but with many systematic
biases and
much variability across the genome. \citet{Campbell2008}
was one of the first to use genome-wide sequencing to detect CNA
events. The reader
is also referred to \citet{Medvedev2009} for a review
of recent studies in CNV/CNA detection using sequencing data.
More details of the data, with an illustrative example (Figure \ref
{figillusReads}),
are given in Section \ref{secDataExistingMethod}.

In the shift from array-based to sequencing-based copy number profiling,
the main statistical challenge arises from the fundamental change
in the type of data observed. Array-based data are represented by
a large but fixed number of continuous valued random variables that are
approximately
normal after appropriate preprocessing, and CNV/CNA signals based on
array data can
be modeled as shifts in mean. Sequencing-based data, as we will discuss
further in Section \ref{secDataExistingMethod}, are realizations of point
processes, where CNV/CNA signals are represented by shifts in intensity
of the process. While one can apply a normal approximation to the large
number of discrete events in sequencing data, hence translating the
problem into
the familiar array-based setting, this approach is inefficient
and imprecise. A more direct model of the point process is preferred.
This type of data calls for a new statistical model, new test
statistics, and, due to the quick growth
of sequencing capacity, new and highly efficient computing implementation.

In copy number profiling it is important to assess the confidence
in the estimated copy numbers. With the exception of \citet{LaiXingZhang},
existing segmentation methods, both for array data and for sequencing
data, give a hard segmentation and do not quantify the uncertainty in their
change-point estimates. Some methods, such as \citet{Olshen2004} and
\citet{Wang2005}, provide confidence
assessments for the called CNV or CNA regions, in the form of false
discovery rates or
$p$-values, thus inherently casting the problem in a hypothesis
testing framework. However, for the analysis of complex regions with
nested changes, such as those in tumor data, confidence intervals
on the copy number, from an estimation perspective, are often more useful.
Intuitively, the copy number estimate is less reliable for a region
near a change point than for a region far away from any change points.
Also, copy number estimates are more reliable for
regions with high coverage than for regions with low coverage, since
coverage directly
affects the number of observations used for estimation.
This latter point makes confidence intervals particularly
important for interpretation of results derived
from short read sequencing data, where coverage can be highly uneven
across the genome.
In this paper, we take a Bayesian approach with noninformative priors
to compute point-wise confidence intervals, as described in Section \ref
{secBayesianCI}.

The proposed methods are based on a simple and flexible inhomogeneous
Poisson Process model for sequenced reads. We derive
the score and generalized likelihood ratio statistics for this model
to detect regions where the read intensity shifts in the target sample,
as compared to a reference. We construct a modified Bayes information
criterion (mBIC) to
select the appropriate number of change points and propose Bayesian
point-wise confidence intervals as a way to assess the confidence in
the copy number estimates. As a proof of concept,
we apply seqCBS, our sequencing-based CNV/CNA detection algorithm,
to a number of actual data sets and found it to have good concordance
with array-based results. We also conduct a spike-in study and compare
the proposed method to SegSeq, a method proposed by \citet{Chiang2009}.

The methods developed in this paper have been implemented in an open-source
R-package, SeqCBS, available from CRAN
\texttt{\href{http://cran.r-project.org/web/packages/seqCBS/index.html}{http://cran.r-project.}
\href{http://cran.r-project.org/web/packages/seqCBS/index.html}{org/web/packages/seqCBS/index.html}}.

\section{Data and existing methods}\label{secDataExistingMethod}
\begin{figure}

\includegraphics{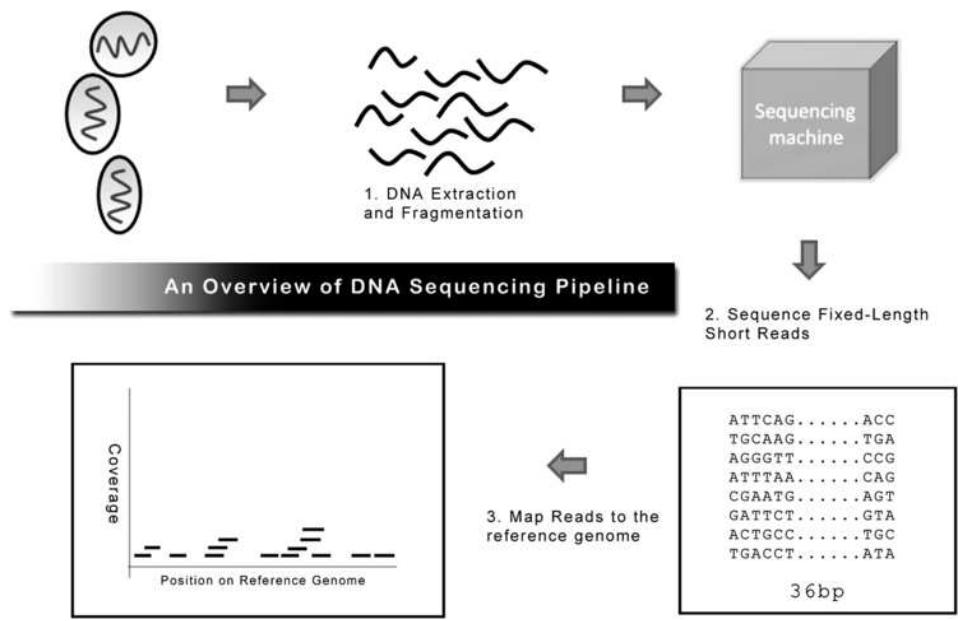}

\caption{Overview of sequencing pipeline.}\label{figseqPipe}
\end{figure}

In a general next-generation genome sequencing/resequencing pipeline,
shown in Figure \ref{figseqPipe}, the DNA in the sample
is randomly fragmented, and a short sequence of the ends of the
fragments is ``read'' by the sequencer.
After the bases in the reads are called, the reads are mapped to
the reference genome. There are many different approaches to the
preparation of the DNA library prior
to the sequencing step, some involving amplification by polymerase
chain reaction, which lead to
different distribution of reads along the reference genome. When a
region of the genome is duplicated,
fragments from this region have a higher representation, and thus its
clones are more likely to be
read by the sequencer. Hence, when mapped to reference genome,\vadjust{\goodbreak} this duplicated
region has a higher read intensity. Similarly, a deletion
manifests as a decrease in read intensity. Since reads are contiguous
fixed length sequences, it suffices
to keep track of the reference mapping location of one of the bases
within the read.
Customarily, the reference mapping location of the 5$'$ end of the read
is stored and
reported. This yields a point process with the reference genome as the
event space.\looseness=-1

As noted in previous studies, sequencing coverage is dependent on
characteristics of the local DNA sequence, and fluctuates even when there
are no changes in copy number, as shown in \citet{Dohm2008}. Just as adjusting
for probe-effects is important for interpretation of microarray data,
adjusting for these baseline fluctuations in depth of coverage is
important for sequencing data. The bottom panels of Figure \ref
{figseqarrayHCC1954}
show the varying depth of coverage for Chromosomes 8 and 11 in the
sequencing of a normal human sample, HCC1954. Many factors cause the
inhomogeneity of depth of coverage. For example, regions of the genome
that contain
more G/C bases are typically more difficult to fragment in
an experiment. This results in lower depth of coverage in such regions.
Some regions of the genome
are highly repetitive. It is challenging to map reads from repetitive
regions correctly
onto the reference genome and, hence, some of the reads are inevitably
discarded as
unmappable, resulting in loss of coverage in that region, even though
no actual deletion has occurred. Some ongoing efforts on the analysis
of sequencing data involve modeling
the effects of measurable quantities, such as GC content and
mappability, on baseline depth.
\citet{Cheung06062011} demonstrated that read counts in sequencing are
highly dependent on
GC content and mappability, and discussed a method to account for such
systematic biases.
\citet{BenjaminiSpeed2011TR} investigated the relationship between GC
content and read count on the
Illumina sequencing platform with a single position model, and
identified a family of unimodal curves
that describes the effect of GC content on read count.
We take the approach of empirically controlling for the baseline fluctuations
by comparing the sample of interest to a control sample that was prepared
and sequenced by the same protocol. In the context of tumor CNA
detection, the control
is preferably a matched normal sample, for it eliminates the discovery
of germline
copy number variants and allows one to focus on somatic CNA regions of
the specific tumor genome.
If a perfectly matched sample is not possible, a~carefully chosen control
or a pool of controls, with sequencing performed on the same platform
with the same experimental protocol, would work for our method as well since
almost all of the normal human genome are identical.

\begin{figure}

\includegraphics{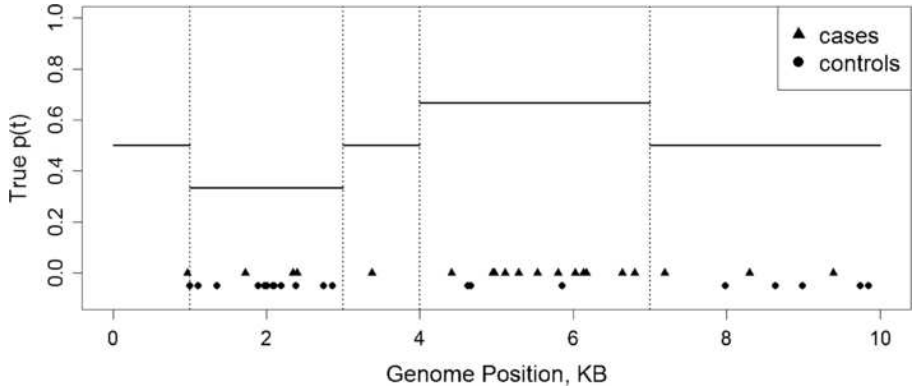}

\caption{Illustration of paired Poisson Processes and $p(t)$.}\label{figillusReads}
\end{figure}

As a simple and illustrative example of the data, we generated points
according to a
nonhomogeneous Poisson Process. Figure \ref{figillusReads} shows the
point processes and the underlying $p(t)$ function, defined as the probability
that a read at genomic position~$t$ is from the case/tumor sample,
conditional on
the existence of a read at position~$t$. The model is discussed in more detail
in Section \ref{secTheModel}. The $y$-values for the points are jittered
for graphical clarity.

Existing methods on CNV and CNA detection with sequencing data generally
follow the change-point paradigm, which is natural since copy number
changes reflect actual breakpoints along chromosomes. \citet{Chiang2009}
proposed the algorithm SegSeq that segments the genomes of a tumor
and a matched normal sample by using a sliding fixed size window,
reducing the data to the ratio of read counts for each window.
\citet{Xie2009} proposed CNV-seq that detects CNV regions
between two individuals based on binning the read counts and then
applying methods developed for array data. \citet{Yoon2009}
designed a method named Event-Wise Testing (EWT) that detects CNV
events using a fixed-window scan on the GC content adjusted read counts.
\citet{Ivakhno15122010} proposed a method called CNAseg that uses read
counts in
windows of predefined size, and discovers CNV using a Hidden Markov Model
segmentation. As for single sample CNV detection method, \citet{Boeva15012011}
constructed a computational algorithm that normalizes read counts by GC content
and estimates the absolute copy number.

These existing methods approach this statistical problem by binning
or imposing fixed local windows. Some methods utilize the log ratio
of read counts in the bin or window as a test statistic, thereby reducing
the data to the familiar representation of array-based CNV/CNA detection,
with \citet{Ivakhno15122010} being an exception in that it uses the difference
in tumor-normal window read counts in their HMM segmentation.
There are a number of downsides to the binning or local window approach.
First, due to the inhomogeneity of reads, certain bins will receive
much larger number of reads overall than other bins, and the optimal
window size varies across the genome. If the number of reads in a
bin is not large enough, the normal approximations that are employed
in many of these methods break down. Second, by binning or fixed-size
window sliding, the estimated CNV/CNA boundaries can be imprecise
if the actual breakpoints are not close to the bin or window boundary.
This problem can be somewhat mitigated by refining the boundary after
the change point is called, as done in SegSeq. In this paper, we
propose a unified model, one that detects the change points, estimates
their locations, and assesses their uncertainties simultaneously.

To illustrate and evaluate our method, we apply it to real and
spiked-in data based on a pair of NCI-60 tumor/normal cell lines,
HCC1954 and BL1954. The data for these samples were produced and
investigated by \citet{Chiang2009}.
The whole-genome shotgun sequencing was performed on the Illumina
platform and the reads are 36~bp long. After read and mapping quality
exclusions, 7.72 million and 6.65 million reads were used for the tumor
(HCC1954) and normal (BL1954) samples, respectively. Newer sequencing
platforms produce much more massive data sets.

\section{A change-point model on two nonhomogeneous Poisson
processes}\label{secTheModel}

We start with a statistical model for the sequenced reads. Let $\{
X_{t}\vert t\leq T\} $
and $\{ Y_{t}\vert t\leq T\} $ be the number of reads
whose first base maps to the left of base location $t$ of a given
chromosome for the case and control samples, respectively. We can
view these count processes as realizations of two nonhomogeneous
Poisson processes (NHPP), one each for the case and control samples,
\begin{eqnarray}
\{ X_{t}\} &\sim&\operatorname{NHPP}(\mu_{t}),
\nonumber
\\[-8pt]
\\[-8pt]
\nonumber
\{ Y_{t}\} &\sim&\operatorname{NHPP}(\lambda_{t}).
\end{eqnarray}
The scale $t$ is in base pairs. The scenario
where two or more reads are mapped to the same genomic position is allowed
by letting $\mu_{t}$ and $\lambda_{t}$ take values larger than~1
and assuming that the observed process is binned at the integers.
We propose a change-point model on the conditional probability of
an event at position $t$ being from $\{ X_{t}\} $, given that
there is such an event from either $\{ X_{t}\} $ or $\{
Y_{t}\} $,
namely,
%
\begin{equation}
p(t)=\frac{\mu_{t}}{\mu_{t}+\lambda_{t}}=p_{k}\qquad\mbox{if
}t_{k}\leq t<t_{k+1}, k=1,\ldots,K.\label{eqModelSpec,ActualScale}
\end{equation}

An example of data according to this model is shown in Figure \ref
{figillusReads}. The change-point model assumption can be equivalently
expressed as

\[
\mu_{t}=\lambda_{t}f(t),
\]
where $f(t)=p(t)/[1-p(t)]$ is piecewise constant with
change points $\{ t_{k}\} $. Of course, we require the
collection of change points to lie within the observation window:
\[
0=t_{0}<t_{1}<\cdots<t_{K+1}=T.
\]

This model does not force the overall intensity of case and control
reads to be the same. The intensity function $\lambda_{t}$ reflects
the inhomogeneity of the control reads. One interpretation of the
model is that, apart from constant shifts, the fluctuation of coverage
in the case sample is the same as that in the control sample. This
is reasonable if the case and control samples are prepared and sequenced
by the same laboratory protocol and mapped by the same procedure, as we
discussed in Section \ref{secDataExistingMethod}. The model
would not be valid if the intensity functions for samples have significant
differences caused by nonmatching protocols or experimental biases.

Let $\{ U_{1},\ldots,U_{m_1}\} $,
$\{ V_{1},\ldots,V_{m_2}\} $ be the event locations for processes
$\{ X_{t}\} $ and $\{ Y_{t}\} $, respectively.
That is, $U$ and $V$ are the mapped positions of the reads from
the case and control samples. Let $m=m_1+m_2$ be the total number
of reads from the case and control samples combined.
We combine the read positions from the case and control processes
and keep them ordered in the genome position, obtaining combined read
positions $W_{1},\ldots,W_{m}$
and indicators of whether each event is a realization of the case
process or the control process $Z_{1},\ldots,Z_{m}$:
%
\begin{equation}\label{eqZiDefinition}
Z_{i}=
\cases{
1, &\quad$\mbox{if }W_{i}\in\{ U_{1},\ldots,U_{m_1}\},$\vspace*{2pt}\cr
0,&\quad $\mbox{if }W_{i}\in\{ V_{1},\ldots,V_{m_2}\}.$
}
\end{equation}
For any read $i$ in the combined process, we will sometimes use the
term ``success'' to mean that $Z_{i}=1$, that is, that the read is
from the case process. Notice that the collection of change-point
locations that can be inferred with the data is precisely $\{
W_{1},\ldots, W_{m}\} $,
since we do not have data points to make inference in favor of or against
any change points in between observations. This means that
estimating the copy number between two genome positions is equivalent to
doing so for the closest pair of reads
that span the two genome positions of interest. Namely, there is
a one-to-one correspondence between the set of possible change
points on $\{ W_{1},\ldots, W_{m}\} $ and the set of change
points $\{ 1=\tau_{0}<\tau_{1}<\cdots<\tau_{K+1}=m\} $
defined on the indices $\{ 1,\ldots,m\} $ through the following:
\[
\{ \tau_{k}=j\} \Longleftrightarrow\{ t_{k}=W_j\}.
\]

The above statement can be made formal through the equivariance
principle. Consider
the sample space $[0,T]$ of any change point, any
monotonically increasing
function $\phi:[0,T]\rightarrow[0,T]$, and its
natural vector extension $\bar{\phi}(c_1,\ldots,c_n)=
(\phi(c_1),\ldots,\phi(c_n))$.

\begin{defn}
A change-point estimator $\hat{\tau}$ is monotone transform equivariant
if for all monotonically
increasing functions $\phi:[0,T]\rightarrow[0,T
]$, we have
$\hat{\tau}(\bar{\phi}(U),\bar{\phi}(V)
)=\bar{\phi}(\hat{\tau}(U,V)).$
\end{defn}

The following theorem shows that any breakpoint estimator $\hat{\tau
}(U,V)$ satisfying
the equivariance condition can be decomposed into a simpler form.

\begin{thmm}\label{thmequivThm}
Let $\hat{\tau}(U,V)$, which takes values in $W$, be an estimator of
the breakpoints. Then $\hat{\tau}$ is monotone
transform equivariant if and only if $\hat{\tau}(U,V)=W_{\hat
{K}}$, where $\hat{K}=f(Z)$
taking integer values $\{ 1,\ldots,m\}$ does not depend on $W$.
\end{thmm}

\begin{pf}
For ease of notation, we let $\bar{\phi}(W)=(\phi
(W_{1}),\ldots,\phi(W_{m}))$
be the natural extension of $\phi$. Suppose $\hat{\tau}(U,V
)=W_{\hat{K}}$,
where $\hat{K}=f(Z)$. Note that~$Z$ is invariant to all
monotone transformations of the arrival times, hence so is $\hat{K}$.
Therefore, $\phi(\hat{\tau}(U,V))=\phi
(W_{\hat{K}})=(\bar{\phi}(W))_{\hat{K}}=\hat
{\tau}(\bar{\phi}(U),\bar{\phi}(V))$.

In the other direction, since $\hat{\tau}\in W$ and $(U,V)$
contain the same information as $(W,Z)$, we must have
$\hat{\tau}(U,V)=W_{\hat{K}(W,Z)}$. Suppose
that $\hat{K}(W,Z)$ depend on $W$ in a nontrivial way
but $\hat{\tau}$ satisfies the monotone transform equivariance condition.
This means that there exist $W'\neq W$ such that $\hat{K}(W,Z
)\neq\hat{K}(W',Z)$.
But since~$W$ and $W'$ are both increasing finite sequences on $
[0,T]$
with the same number of elements, we must have some $\phi(\cdot
)$
that $\bar{\phi}(W)=W'$. Note that $(W',Z)$
induces $(U',V')=(\bar{\phi}(U),\bar{\phi
}(V))$.
However, $\hat{\tau}(\bar{\phi}(U),\bar{\phi}
(V))=\hat{\tau}(U',V')=W'_{\hat{K}
(W',Z)}=\phi(W_{\hat{K}(W',Z)})\neq\phi
(W_{\hat{K}(W,Z)})=\break\phi(\hat{\tau}(U,V
))$.
Hence, the equivariance property holds if and only if $\hat{K}$ is only
a function of $Z$.
\end{pf}

Theorem \ref{thmequivThm} implies that any breakpoint estimation
procedure, that is, monotone transform equivariant uses
the estimator $\hat{K}$ of integer breakpoints based on $Z$, and that
the actual read position $W$ merely serves as a genomic scale lookup table.
Hence, we can define our change-point
model on the indices $\{ 1,\ldots,m\} $ for the read counts,
and use the conditional likelihood which depends only on $\{
Z_{i}\} $
but not on the event positions $\{ W_{1},\ldots,W_{m}\} $:
%
\begin{equation}
p(j)=p_{k}\qquad\mbox{if }\tau_{k}\leq j<\tau_{k+1}.\label{eqModelSpec,indexScale}
\end{equation}

For the rest of this section, we will exclusively work with equation
(\ref{eqModelSpec,indexScale}). The mapping positions $\{
W_{1},\ldots,W_{m}\} $
will re-enter our analysis when we compute confidence intervals for
the copy number estimates, in Section \ref{secBayesianCI}. Our statistical
problem is hence two-fold. First, given $K$, we need to estimate
the change points $\{ \tau_{k}\}$. Second, we need a method
to select model complexity, as dictated by $K$.

We start by considering the following simplified problem: Given a
single interval spanning reads $i$ to $j$ in the combined process,
we want to test whether the success probability inside this interval,
$p_{ij}$, is different from the overall success probability, $p$.
The null model $H_0$ states that $p_{ij}=p$. We derive two statistics
to test this hypothesis. The first is adopted from the conditional
score statistic for a general exponential family model where the signal
is represented by a kernel function, as discussed in \citet
{RabinowitzPointScanStat1994},
%
\begin{equation}
\qquad S_{ij}=\sum_{k=1}^{m}(Z_{k}-\hat{p})\Biggl(1_{i\leq k\leq
j}-\frac{1}{m}\sum_{k=1}^{m}1_{i\leq k\leq j}\Biggr)=\sum_{i\leq k\leq
j}Z_{k}-\hat{p}(j-i+1),
\end{equation}
where $\hat{p}=\sum{Z_k} / m$. This statistic is
simply the difference between the number of observed
and expected case events under the null model. Its variance at the null is
\[
\hat{\sigma}_{ij}^{2} = \operatorname{Var}(S_{ij}) = \biggl(1-\frac
{j-i+1}{m}\biggr)(j-i+1)p(1-p)
\]
and is used to standardize $S_{ij}$
for comparison between regions of different sizes.
The standardized score statistic $T_{ij}=S_{ij}/\sigma_{ij}$
is intuitive and simple, and would be approximately standard normal
if $j-i$ were large. However, the normal
approximation is not accurate if the number of reads that map to the
region is low. To attain higher accuracy for regions with low read
count, observe that $\sum_{i\leq k\leq j}Z_{k}$ is a binomial
random variable, and use an exact binomial generalized likelihood
ratio (GLR) statistic,
\[
\Lambda_{ij}=\sup_{p_{0},p_{ij}}l_{1}(p_{0},p_{ij})-\sup_{p}l_{0}(p),
\]
where the null model with one overall success probability parameter
$p$ is compared with the alternative model with one parameter $p_{ij}$
for inside the $[i,j]$ interval and another parameter
$p_{0}$ for outside the interval.\vadjust{\goodbreak} From the binomial log-likelihood
function one obtains
\begin{eqnarray*}
\Lambda_{ij} &=& \sum_{k\in[i,j]}\biggl\{ Z_{k}\log\biggl(\frac
{\hat{p}_{ij}}{\hat{p}}\biggr)+(1-Z_{k})\log\biggl(\frac
{1-\hat{p}_{ij}}{1-\hat{p}}\biggr)\biggr\}\\
& &{}+\sum_{k\notin[i,j]}\biggl\{ Z_{k}\log\biggl(\frac{\hat
{p}_{0}}{\hat{p}}\biggr)+(1-Z_{k})\log\biggl(\frac{1-\hat
{p}_{0}}{1-\hat{p}}\biggr)\biggr\} ,\label{eqBinomGLR}
\end{eqnarray*}
where $\hat{p}$, $\hat{p}_{0}$, $\hat{p}_{ij}$ are maximum likelihood estimates
of success probabilities
\begin{eqnarray*}
\hat{p}&=&\sum_{k=1}^{m} Z_{k}/m,\\
\hat{p}_{ij}&=&\sum_{k\in[i,j]}Z_{k}/(j-i+1),
\\
\hat{p}_{0}&=&\sum_{k\notin[i,j]}Z_{k}/(m-j+i-1).
\end{eqnarray*}

The GLR and score statistics allow us to measure how distinct a specific
interval $[i,j]$ is compared to the entire chromosome. For the
more general problem in which $(i,j)$ is not given but
only one such pair exists, we compute the statistic for all unique
pairs of $(i,j)$ to find the most significantly distinct
interval. This operation is $O(m^{2})$ and to improve
efficiency, we have implemented a search-refinement scheme called
Iterative Grid Scan in our software. It works by identifying larger
interesting intervals on a coarse grid and then iteratively improving
the interval boundary estimates. The
computational complexity is roughly $O(m\log m)$ and hence
scales easily. A similar idea was studied in \citet{Walther2010}.

In the general model with multiple unknown change points,
one could theoretically estimate all change points
simultaneously by searching through all possible combinations of $
\{ \hat{\tau}_{k}\} $.
But this is a combinatorial problem where even the best dynamic programming
solution [\citet{Bellman1961}; \citet{BaiPerron2003}; \citet{Lavielle2005}] would not scale
well for a data set containing millions of reads. Thus, we adapted
Circular Binary Segmentation [\citet{Olshen2004}; \citet{Venkatraman2007}] to
our change-point model as a greedy alternative. In short, we find the
most significant region $(i,j)$ over the entire chromosome,
which divides the chromosome in to 3 regions (or two, if one of the
change points lies on the edge). Then we further scan each of the
regions, yielding a candidate subinterval in each region. At each
step, we add the most significant change point(s) over all of the
regions to the
collection of change-point calls.

Model complexity grows as we introduce more change points.
This brings us to the issue of model selection: We need a method to
choose an appropriate number of change points $K$. \citet{Zhang2007}
proposed a solution
to this problem for Gaussian change-point models with shifts in mean.
Like the Gaussian model, the Poisson change-point model has irregularities
that make classic measures such as the AIC and the BIC inappropriate.
An extension of \citet{Zhang2007} gives a Modified Bayes Information
Criterion (mBIC) for our model, derived as a large sample approximation
to the Bayes Factor in the spirit of \citet{Schwarz1978}:
\begin{eqnarray*}
\operatorname{mBIC}(K) &=& \log\biggl(\frac{\sup_{p(t),\tau}L(p(t),\tau)}{\sup
_{p}L(p)}\biggr)-\frac{1}{2}\sum_{k=0}^{K}\log(\hat{\tau}_{k+1}-\hat
{\tau}_{k})\\
& &{}+\frac{1}{2}\log(m)-K\log(m'),
\end{eqnarray*}
where $m'$ is the number of unique values in $\{ W_{1},\ldots
,W_{m}\} $.

The first term of mBIC is the generalized log-likelihood ratio for the
model with $K$ change points versus the null model with no change
points. In our context,
$K$ ideally reflects the number of biological breakpoints that yield
the copy number variants.
The remaining terms can be interpreted as a ``penalty'' for model
complexity. These penalty terms differ from the penalty term in the
classic BIC of \citet{Schwarz1978} due to nondifferentiability of the
likelihood function
in the change-point parameters $\{ \tau_{k}\},$ and also
due to the fact that the range of values for $\{ \tau_{k}\}$ grow
with the number of observations $m$. For more details on the interpretation
of the terms in the mBIC, see \citet{Zhang2007}.
Finally, we report the segmentation with $\widehat{{K}}=
{\arg\max}_{K}\operatorname{mBIC}(K)$
change points.

\section{Approximate Bayesian confidence intervals}\label{secBayesianCI}

As noted in the \hyperref[sec1]{Introduc-}\break\hyperref[sec1]{tion}, it is particularly important for
sequencing data to assess the uncertainty
in the relative read intensity function at each genomic position.
We approach this problem by constructing approximate Bayesian confidence
intervals.

Suppose $Z_{1},\ldots,Z_{m}$ are independent realizations of Bernoulli
random variables with success probabilities $\{p_{t}\}$. Consider first
the one change-point model (which can be seen as a local part of
a multiple change-point model), where
\[
p_{t}=
\cases{
p_{0}, & \quad$\mbox{if }t\leq\tau,$\vspace*{2pt}\cr
p_{1}, & \quad$\mbox{if }t>\tau.$
}
\]
Without loss of generality, we may take $\tau\in\{ 1,2,\ldots
,m\} $.
Assume a uniform prior for $\tau$ on this discrete set. Let $S_{t}$
be the number of successes up to and including the $t${th} realization,
\[
S_{t}=\sum_{1\leq i\leq t}Z_{i}.
\]

Our goal is to construct confidence bands for $p_{t}$ at each
$t\in\{ 1,2,\ldots,m\} $. Assume a $\operatorname{Beta}(\alpha
,\beta)$
prior for $p_{0}$ and $p_{1}$. If we knew $\tau$, then the
posterior distribution of $p_{0}$ and $p_{1}$ is
\begin{eqnarray*}
f(p_{0}\vert\vec{Z}, \tau=\tau^*) & \propto& f
(p_{0})f(S_{\tau^{*}}\vert p_{0})\\
& \sim& \operatorname{Beta}(\alpha,\beta)\cdot\operatorname{Binom}
(S_{\tau^{*}}; \tau^{*},p_{0})\\
& \sim& \operatorname{Beta}(\alpha+S_{\tau^{*}},\beta+\tau^{*}-S_{\tau
^{*}}),
\\
f(p_{1}\vert\vec{Z}, \tau=\tau^*) & \propto& f
(p_{1})f(S_{m}-S_{\tau^{*}}\vert p_{1})\\
& \sim& \operatorname{Beta}(\alpha+S_{m}-S_{\tau^{*}},\beta+m-\tau
^{*}-S_{m}+S_{\tau^{*}}).
\end{eqnarray*}
Now, without knowing the actual $\tau^{*}$, we compute the posterior
distribution of $p_{t}$ as
\begin{eqnarray*}
f(p_{t}\vert\vec{Z}) & = & \sum_{i=1}^{m}f(p_{t},\tau
=i\vert\vec{Z})\\
& = & \sum_{i=1}^{m}f(p_{t}\vert\tau=i,\vec{Z})\cdot f
(\tau=i\vert\vec{Z}).
\end{eqnarray*}
As before, the first part of the summation term is a beta distribution,
\[
f(p_{t}\vert\tau=i,\vec{Z})=
\cases{
\operatorname{Beta}(\alpha+S_{i},\beta+i-S_{i}), & \quad $\mbox{if }t\leq i,$\vspace*{2pt}\cr
\operatorname{Beta}(\alpha+S_{m}-S_{i},\beta+m-i-S_{m}+S_{i}), & \quad $\mbox{if
}t>i,$}
\]
and for the second term, we define the likelihood of the change point
at $i$ as $L_{i}=f(\vec{Z}\vert\tau=i)$
and observe that with the uniform prior on $\tau$,
\[
f(\tau=i\vert\vec{Z})\propto L_{i}/m\propto L_{i},
\]
where
%
\begin{equation}
L_{i}=\int\prod_{1\leq j\leq i}p_{0}^{Z_{j}}(1-p_{0}
)^{1-Z_{j}}\,dP_{0}\cdot\int\prod_{i<j\leq m}p_{1}^{Z_{j}}
(1-p_{1})^{1-Z_{j}}\,dP_{1},\label{eqBayesLikDef}
\end{equation}
and $dP_{0}$ and $dP_{1}$ are with respect to the prior distributions
of $p_{0}$ and $p_{1}$. With $\operatorname{Beta}(\alpha, \beta)$ priors on
$p_{0}$ and $p_{1}$, we can find the closed form expression of $L_i$:
%
\begin{eqnarray}\label{eqBayesLikForm}
L_{i} & = & \int\prod_{1\leq j\leq i}p_{0}^{Z_{j}}(1-p_{0}
)^{1-Z_{j}}\,dP_{0}\cdot\int\prod_{ i<j\leq m}p_{1}^{Z_{j}}
(1-p_{1})^{1-Z_{j}}\,dP_{1}\nonumber\\
& = & \frac{1}{B(\alpha,\beta)}\int p_{0}^{S_{ i}}
(1-p_{0})^{ i-S_{ i}}p_{0}^{\alpha-1}(1-p_{0})^{\beta
-1}\,dp_{0}\nonumber\\
& &{} \times\frac{1}{B(\alpha,\beta)}\int p_{1}^{S_{m}-S_{
i}}(1-p_{1})^{m- i-S_{m}+S_{ i}}p_{1}^{\alpha-1}
(1-p_{1})^{\beta-1}\,dp_{1}
\nonumber
\\[-8pt]
\\[-8pt]
\nonumber
& = & \frac{B(\alpha+S_{ i},\beta+ i-S_{ i})B(\alpha
+S_{m}-S_{ i},\beta+m- i-S_{m}+S_{ i})}{B(\alpha,\beta
)^{2}}\\
& = & \frac{\Gamma(\alpha+S_{ i})\Gamma(\beta+ i-S_{
i})}{\Gamma(\alpha+\beta+ i)}\nonumber\\
& & {}\times\frac{\Gamma(\alpha+S_{m}-S_{ i})\Gamma(\beta+m-
i-S_{m}+S_{ i})}{\Gamma(\alpha+\beta+m- i)}\frac{\Gamma
(\alpha+\beta)^{2}}{\Gamma(\alpha)^{2}\Gamma
(\beta)^{2}}.\nonumber
\end{eqnarray}
Hence, we can compute, without knowing the actual value of $\tau$,
%
\begin{equation}\label{newlab}
f(p_{t}\vert\vec{Z})  \propto \sum_{i=1}^{m}f
(p_{t}\vert\tau=i,\vec{Z})\cdot\frac{L_{i}}{L_{\hat{\tau}}}
\qquad \mbox{where }\hat{\tau}=\arg\max_{i}L_{i}.
\end{equation}

Observe that the posterior distribution is a mixture of $\operatorname
{Beta}(\cdot,\cdot)$
distributions. In theory, we could compute weights $w_{i}=L_{i}/L_{\hat
{\tau}}$
for all positions~$i$ and then numerically compute $(\frac{\alpha
}{2},1-\frac{\alpha}{2})$
quantiles of the posterior beta mixture distribution to obtain the
Bayesian confidence intervals. However, in practice, one can approximate
the sum in (\ref{newlab}) by
\[
f(p_{t}\vert\vec{Z})\approx\frac{1}{\sum_{w_{i}>\varepsilon
}w_{i}}\biggl[\sum_{w_{i}>\varepsilon}w_{i}f(p_{t}\vert\tau=i,\vec
{Z})\biggr],
\]
for some small $\varepsilon>0$, hence ignoring the highly unlikely locations
for the change points. Empirically, we use $\varepsilon=10^{-4}$.
It is easy to see that the sequence of log likelihood ratios for
alternative change points,
$\log{\frac{L_i}{L_{\tau}}}$, form random walks with negative drift as
$i$ moves
away from the true change point $\tau$ [\citet{HINKLEY01041970}]. The
negative drift depends on the true $p_0$, $p_1$
and is larger in absolute magnitude when the difference between $p_{0}$
and $p_{1}$ is larger. With
$\tau$ unknown, since $P(\vert\hat{\tau} - \tau\vert\leq\delta
)$ can be made arbitrarily close to 1 for $\delta=o(m)$, one can
make the same random walk construction for $\log(L_i/L_{\hat{\tau}})$
bounded away by $\delta$ from $\hat{\tau}$, as done in \citet{COBB01081978}.
This implies that, for any $\varepsilon>0$,
one may find a constant $c_{\varepsilon,p_{0},p_{1}}$ such that for any~$i$
at least $c_{\varepsilon,p_{0},p_{1}}$ steps away from $\hat{\tau}$,
$w_{i}<\varepsilon$ with probability approaching~1. Hence, it
is reasonable to use a small cutoff to produce a close approximation
to the posterior distribution.

The extension of this construction to multiple change points is straightforward.
It entails augmenting the mixture components of one change point with
that of its neighboring change points. This gives a computationally
efficient way of approximating the Bayesian confidence interval using,
typically, a few hundred mixture components, which has been implemented
in \texttt{seqCBS}. There is also an extensive body of literature on
constructing
confidence intervals and confidence sets for estimators of the change
point $\tau$.
We refer interested readers to \citet{Siegmund1988} for discussion and efficiency
comparison of various confidence sets in change-point
problems.

\section{Results}

We first applied the proposed method to a matched pair of tumor and
normal NCI-60 cell lines, HCC1954 and BL1954. \citet{Chiang2009}
conducted the sequencing of these samples using the Illumina platform.
For comparison with array-based copy number profiles
on the same samples, we obtained array data on HCC1954 and BL1954
from the NCI-60 database at \url{http://www.sanger.ac.uk/genetics/CGP/NCI60/}.
We applied the CBS algorithm [\citet{Olshen2004}, \citet{Venkatraman2007}] with
modified BIC stopping algorithm [\citet{Zhang2007}] to estimate
relative copy numbers based on the array data.

%
\begin{figure}
\centering
\begin{tabular}{@{}cc@{}}

\includegraphics{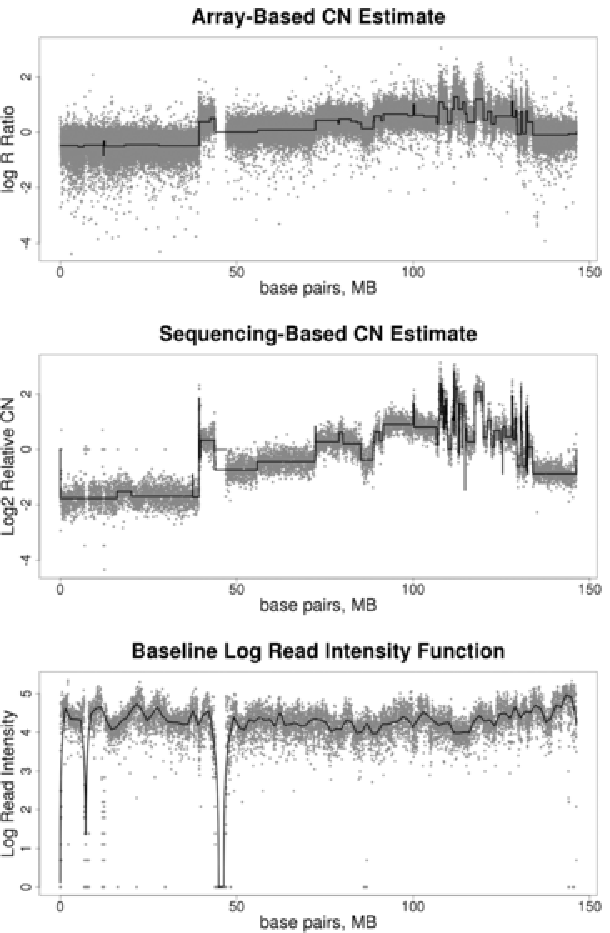}
 & \includegraphics{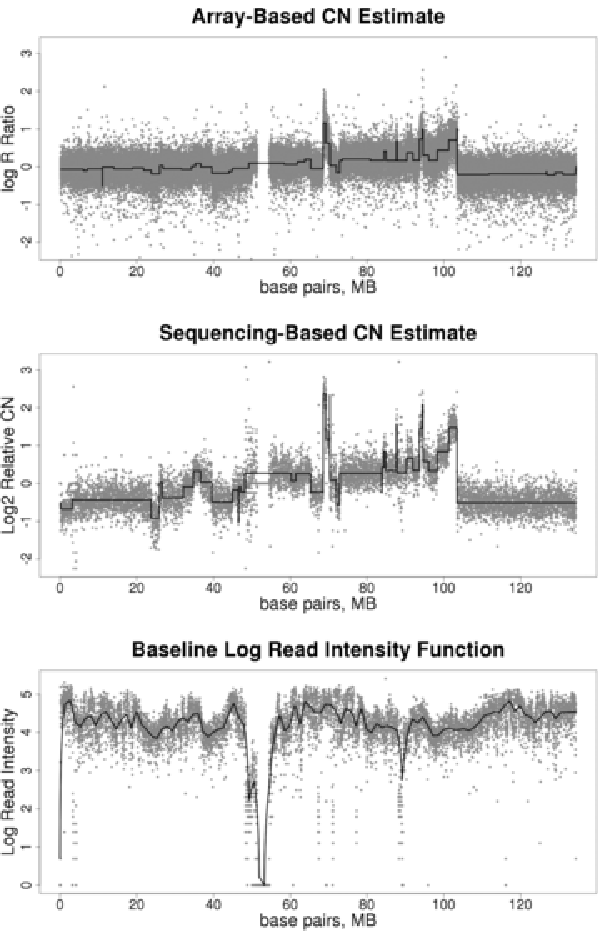}\\
\footnotesize{(a)} & \footnotesize{(b)}
\end{tabular}
\caption{Comparison of seqCBS and array-based CN profiling.
\textup{(a)} HCC/BL1954 Chr 8, array, seqCBS, baseline read intensity.
\textup{(b)} HCC/BL1954 Chr 11, array, seqCBS, baseline read intensity.}\label{figseqarrayHCC1954}
\end{figure}

Figure \ref{figseqarrayHCC1954} shows the copy number profiles
estimated from the array data (top) and from the sequencing data
by seqCBS (middle), for two representative chromosomes where there
appears to be a number of copy number alternation events. The bottom
plots show the baseline $\lambda(t)$ function estimated
by smoothing the binned counts of the normal sample sequencing data.
There is clearly inhomogeneity in the rate function. The points
for the top plots are normalized $\log$ ratios for the intensities
of each probe on the array, whereas those for the middle plots are
log ratios of binned counts for the tumor and normal samples. Note
that the binned counts for the sequencing data are only for illustrative
purposes, as the proposed method operates on the point processes directly,
which are difficult to visualize at the whole-chromosome scale. The
piecewise constant lines indicate the estimated $\log$ relative copy
numbers and the change-point locations. Note that after adjusting
for overall number of read differences between the two samples, the
relative copy number is estimated by $\hat{p}(t)/(1-\hat
{p}(t))$,
where $\hat{p}(t)$ is the MLE estimate of the success
probability of the segment into which $t$ falls.

The shape of the profile and overall locations for most change-point calls
are common between the array and sequencing data. That is, CBS and
SeqCBS applied on data generated from two distinct platforms generally
agree. It is interesting that in the regions where a large
number of CNA events seem to have occurred, our proposed method with
sequencing is able to identify shorter and more pronounced CNA events.
It also appears that the CNA calls based on sequencing are smoother
in the sense that small magnitude shifts in array-based results, such as
the change points after 102~Mb of Chromosome 11, are ignored by seqCBS.
Similar observations can be made with results
on other chromosomes as well. Since we do not know the ground truth in these
tumor samples, it is hard to assess in more detail the performance of the
estimates. Detailed spike-in simulation studies in the next section
give a
systematic view of the accuracy of the proposed method, as compared to
the current standard approach.

\begin{figure}

\includegraphics{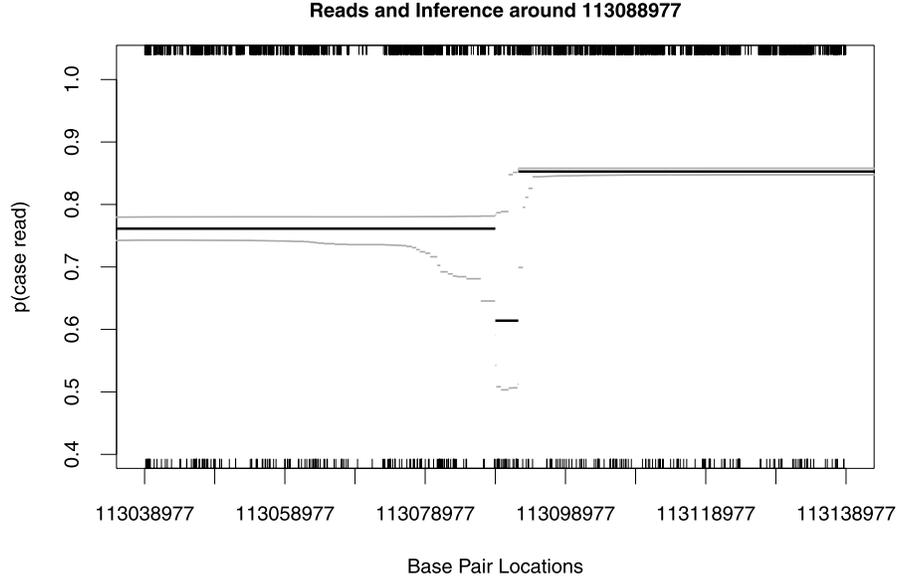}

\caption{Bayesian CI, HCC1954 Chr 8. The solid black line is the
estimated copy number, light gray lines are 95\% confidence intervals.
The actual locations of the case and control reads are shown as tick
marks at the top and bottom of the plot, respectively.}
\label{figbayesCIExample}
\end{figure}

Figure \ref{figbayesCIExample} is an example illustrating the approximate
Bayesian confidence interval estimates around two CNA events on Chromosome
8 of HCC1954. This is not a typical example among the change-point
calls. Typically, the signal differences are stronger and, hence,
the confidence bands are narrower with little ambiguity region.
The actual locations of mapped reads are shown as
tick marks, with ticks at the bottom of the plot representing control
reads, and
ticks at the top representing case reads. The estimated relative copy numbers
and their point-wise approximate Bayesian confidence intervals are
shown as black and grey lines, respectively. One can see that the
width of the confidence intervals depends not only on the number of
reads in the segment, but also on the distance from the position of
interest to the called change points, and that the confidence
intervals are not necessarily symmetric around the estimated copy number.

%

\section{Performance assessment}

To assess the performance of the proposed method more precisely, we
conducted a spike-in simulation study. We empirically estimated the
underlying inhomogeneous rate function $\lambda(t)$ by
kernel smoothing of the read counts from the normal sample, BL1954,
in \citet{Chiang2009}. The simulated tumor rate function
$\mu(t)$ is then constructed by spiking into $\lambda
(t)$
segments of single copy gain/loss. Since the length of the CNA events
influences their ease of detection, we considered a range of different
signal lengths. Each simulated case sample contains 50 changed
segments. We compared
seqCBS to SegSeq, which is one of the more popular available
algorithms. For seqCBS, we used
mBIC to determine appropriate model complexity. We used SegSeq with
its default parameters. A change-point call was deemed true if it was
within 100 reads of a true spike-in change point, after using
a matching algorithm implemented in the R package \texttt{clue}
by \citeauthor{Hornik2010} (\citeyear{Hornik2010,Hornik2010R}) to find minimal-distance pairing
between the called change points and the true change points.
Performance was evaluated by recall and precision, defined respectively
as the
proportion of true signals called by the method and the proportion
of signals called that are true. The simulation was repeated multiple
times to reduce the variance in
the performance measures.

%
\begin{figure}

\includegraphics{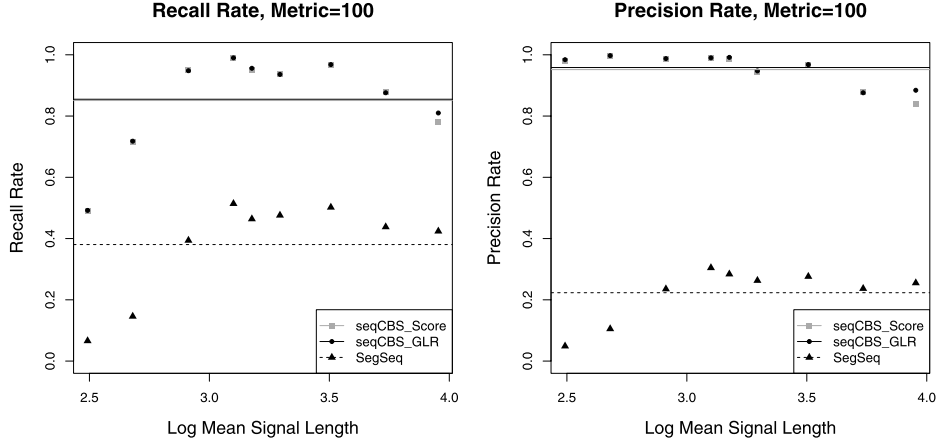}

\caption{Recall and precision of seqCBS \& SegSeq.}\label{figperfOverall}
\end{figure}

\begin{figure}[b]
\centering
\begin{tabular}{@{}c@{\hspace*{2pt}}c@{}}

\includegraphics{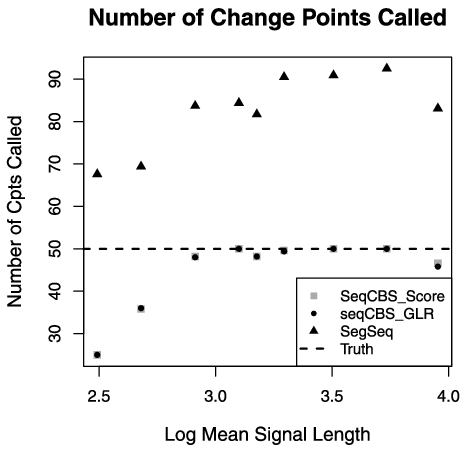}
 & \includegraphics{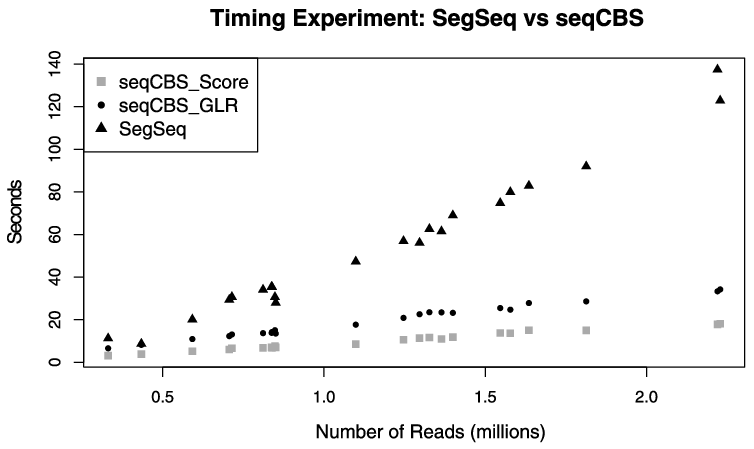}\\
\footnotesize{(a)} & \footnotesize{(b)}
\end{tabular}
\caption{Model complexity and timing by seqCBS \& SegSeq.
\textup{(a)} Model complexity comparison.
\textup{(b)}~Timing comparison.}\label{fig6}
\end{figure}

Figure \ref{figperfOverall} summarizes the performance comparison at
default settings
for a number of spike-in signal lengths. The horizontal lines are mean
recall and precision rates for the methods. We see that SeqCBS, used
with either the score test statistic or the GLR statistic, offers
significant improvement over the existing method in both precision
and recall. The performances of the score and GLR statistics are very similar,
as their recall and precision curves almost overlap. The improvement
in precision can be largely attributed
to the fact that mBIC provides a good estimate of model complexity,
as can be seen in Figure \ref{fig6}(a).

\begin{table}
\caption{Performance measures and tuning parameters. SegSeq: W${}={}$fixed
local window size (default~500),
A${}={}$number of false positive candidates for initialization (default~1000),
B${}={}$number of false positive segments for termination (default 10).
SeqCBS: Scr${}={}$score statistic, Bin${}={}$GLR statistic, G${}={}$IGS power step size
(default 10); N/A values indicate program failures}\label{tabPerformanceMeasures}
\begin{tabular*}{\textwidth}{@{\extracolsep{\fill}}ld{1.3}d{1.3}d{1.3}d{1.3}d{1.3}d{1.3}d{1.3}d{1.3}d{1.3}@{}}
\hline
\multicolumn{10}{c}{Log mean breakpoint length}\\
Length & 2.49 & 2.68 & 2.91 & 3.10 & 3.18 & 3.29 & 3.51 & 3.73 &
3.95\\[6pt]
\multicolumn{10}{c}{Recall}\\
SegSeq & \multicolumn{9}{c}{} \\
Default & 0.066 & 0.146 & 0.394 & 0.514 & 0.464 & 0.476 & 0.502 & 0.438
& 0.424\\
W250 & 0.23 & 0.422 & 0.67 & 0.662 & 0.59 & 0.654 & 0.64 & 0.564 &
0.516\\
W750 & \multicolumn{1}{c}{NA} & \multicolumn{1}{c}{NA} & 0.148 & 0.206 & 0.248 & 0.36 & 0.408 & 0.384 & 0.346\\
A500 & \multicolumn{1}{c}{NA} & 0.148 & 0.394 & 0.514 & 0.464 & 0.476 & 0.502 & 0.438 &
0.424\\
A2000 & \multicolumn{1}{c}{NA} & 0.146 & 0.394 & 0.514 & 0.464 & 0.476 & 0.502 & 0.438 &
0.424\\
B25 & \multicolumn{1}{c}{NA} & 0.182 & 0.404 & 0.532 & 0.484 & 0.502 & 0.506 & 0.452 &
0.458\\
B5 & \multicolumn{1}{c}{NA} & 0.126 & 0.382 & 0.476 & 0.432 & 0.442 & 0.476 & 0.428 &
0.364\\[3pt]
SeqCBS & \multicolumn{9}{c}{} \\
Scr-Def & 0.49 & 0.714 & 0.95 & 0.988 & 0.95 & 0.936 & 0.968 & 0.878 &
0.782\\
Bin-Def & 0.492 & 0.718 & 0.948 & 0.99 & 0.956 & 0.936 & 0.968 & 0.876
& 0.81\\
Scr-G5 & 0.496 & 0.71 & 0.922 & 0.99 & 0.956 & 0.956 & 0.978 & 0.922 &
0.844\\
Bin-G5 & 0.496 & 0.712 & 0.928 & 0.99 & 0.958 & 0.962 & 0.98 & 0.946 &
0.844\\
Scr-G15 & 0.494 & 0.708 & 0.926 & 0.974 & 0.942 & 0.938 & 0.968 & 0.89
& 0.736\\
Bin-G15 & 0.496 & 0.716 & 0.93 & 0.976 & 0.946 & 0.96 & 0.972 & 0.91 &
0.748\\[6pt]
\multicolumn{10}{c}{Precision}\\
SegSeq & \multicolumn{9}{c}{} \\
Default & 0.049 & 0.105 & 0.235 & 0.305 & 0.284 & 0.263 & 0.276 & 0.237
& 0.255\\
W250 & 0.174 & 0.317 & 0.490 & 0.472 & 0.467 & 0.478 & 0.442 & 0.405 &
0.399\\
W750 & \multicolumn{1}{c}{NA} & \multicolumn{1}{c}{NA} & 0.107 & 0.137 & 0.165 & 0.227 & 0.242 & 0.232 &
0.212\\
A500 & \multicolumn{1}{c}{NA} & 0.097 & 0.235 & 0.305 & 0.284 & 0.263 & 0.276 & 0.237 &
0.255\\
A2000 & \multicolumn{1}{c}{NA} & 0.101 & 0.235 & 0.305 & 0.284 & 0.263 & 0.276 & 0.237 &
0.255\\
B25 & \multicolumn{1}{c}{NA} & 0.101 & 0.203 & 0.278 & 0.254 & 0.243 & 0.246 & 0.219 &
0.240\\
B5 & \multicolumn{1}{c}{NA} & 0.104 & 0.278 & 0.361 & 0.323 & 0.308 & 0.327 & 0.295 &
0.271\\[3pt]
SeqCBS & \multicolumn{9}{c}{} \\
Scr-Def & 0.980 & 0.997 & 0.985 & 0.988 & 0.985 & 0.944 & 0.968 & 0.878
& 0.839\\
Bin-Def & 0.984 & 0.997 & 0.988 & 0.990 & 0.992 & 0.947 & 0.968 & 0.876
& 0.884\\
Scr-G5 & 0.984 & 0.997 & 0.956 & 0.990 & 0.992 & 0.980 & 0.994 & 0.945
& 0.942\\
Bin-G5 & 0.984 & 0.994 & 0.959 & 0.990 & 0.994 & 0.990 & 0.996 & 0.977
& 0.942\\
Scr-G15 & 0.980 & 0.994 & 0.953 & 0.944 & 0.961 & 0.949 & 0.964 & 0.876
& 0.710\\
Bin-G15 & 0.984 & 0.994 & 0.953 & 0.946 & 0.973 & 0.984 & 0.972 & 0.910
& 0.733 \\
\hline
\end{tabular*}
\vspace*{-6pt}
\end{table}

We studied the performance sensitivity on tuning parameters. SegSeq allows
three tuning parameters: local window size (W), number of false
positive candidates for initialization (A),
and number of false positive segments for termination~(B). The proposed
method has a step size parameter (G) that
controls the trade-off between speed and accuracy in our Iterative Grid
Scan component, and hence influences
performance. We varied these parameters and recorded the performance
measures in Table \ref{tabPerformanceMeasures}.
It appears that local window size (W) is an important tuning parameter
for SegSeq, and in scenarios with relatively
short signal length, a smaller W${}={}$250 provides significant improvement
in its performance. This echoes with our
previous discussion that methods using a single fixed window size would
perform less well when the signals are not of
the corresponding length. Some of the parameter combinations for SegSeq
result in program running errors in some scenarios, and
are marked as NA. The step size parameter~(G) in SeqCBS, in constrast,
controls the rate at which coarse segment
candidates are refined and the rate at which the program descends into
searching smaller local change points, rather
than defining a fixed window size. A smaller step size typically yields
slightly better performance. However, the proposed
method is not nearly as sensitive to its tuning parameters. We also conducted
a timing experiment to provide the reader with a sense of the required
computational resources to derive the solution. Our proposed method
compares favorably
with SegSeq as seen in Figure \ref{fig6}(b).
The GLR statistic is slightly more complex to compute than the score
statistic, as is reflected in the timing experiment. However, copy number
profiling is inherently a highly parallelizable computing problem: one
may distribute
the task for each chromosome among a multi-CPU computing grid, hence
dramatically reducing the amount of time required for this analysis.

\section{Discussion}

We proposed an approach based on nonhomogeneous Poisson Processes to
directly model next-generation
DNA sequencing data, and formulated a change-point model to conduct
copy number
profiling. The model yields simple score and generalized likelihood
ratio statistics, as well as a modified Bayes information criterion for
model selection. The proposed method has been applied to real
sequencing data and its performance compares favorably to an existing
method in a spike-in simulation
study.

Statistical inference, in the form of confidence estimates, is very
important for sequencing-based data, since, unlike arrays, the
effective sample size (i.e., coverage) for estimating copy number
varies substantially across the genome. In this paper, we derived a
procedure to compute Bayesian confidence intervals on the estimated
copy number. Other types of inference, such as $p$-values or confidence
intervals on the estimated change points, may also be useful. \citet
{Siegmund1988} compares different types of confidence intervals on the
change points, and the methods there can be directly applied to this
problem. The reader is referred to \citet{RabinowitzPointScanStat1994}
and \citet{Siegmund1988a} for existing methods on significance evaluation.

Some sequencing experiments produce paired end reads, where two short
reads are performed on the two ends of a longer fragment of DNA. The
pairing information
can be quite useful in the profiling of structural genomic changes. It
will be important to extend the approach in this paper to handle this
more complex data type.

A limitation of the proposed method and the existing methods is that
they do not handle allele-specific copy number variants. It is possible
to extend our model to accommodate
this need. With deep sequencing, one may assess whether each loci in a
CNV is heterozygous, and estimate the degree to which each allele
contributes to the gain or loss of copy number, by
considering the number of reads covering the locus with the major
allele versus those with the minor allele. This is particularly helpful
for detecting deletion. Furthermore,
in the context of assessing the allele-specific copy number, existing
SNP arrays have the advantage that the assay targets specific sites for
that problem, whereas to obtain
sufficient evidence of allele-specific copy number variants with
sequencing, a much greater coverage would be required since the
overwhelming majority of reads would land in nonallelic genomic
regions. Spatial models that borrow information across adjacent variant
sites, such as \citet{ChenXingZhang2011} and \citet{OlshenPSCBS2011},
would be helpful for improving power.

Recently, there has been increased attention to the problem of
simultaneous segmentation of multiple samples [\citet
{Lipson2006}; \citet{Shah2007}; \citet{Zhang2010b}; \citet{SiegmundYakirZhang}].
One may also wish to extend this method to the multi-sample setting,
where in addition to modeling challenges, one also needs to address
more sources
of systematic biases, such as batch effects and carry-over problems.

Computational challenges remain in this field. With sequencing capacity
growing at record speed,
even basic operations on the data set are resource-consuming. It is
pertinent to develop
faster and more parallelizable solutions to the copy number profiling problem.

\section*{Acknowledgments}
We thank H.~P. Ji, G. Walther and D.~O. Siegmund for their inputs.

%


\printaddresses

\end{document}